\documentstyle[sprocl]{article}

\def\Journal#1#2#3#4{{#1} {\bf #2}, #3 (#4)}

\def\PLB{{\em Phys. Lett.}  B}

\def\PRD{{\em Phys. Rev.} D}

\def\be{\begin{equation}}
\def\ee{\end{equation}}
\def\bea{\begin{eqnarray}}
\def\eea{\end{eqnarray}}

\begin{document}

\title{WEAK FORM FACTORS FOR HEAVY MESON DECAYS\footnote{Talk presented by D. Melikhov}}

\author{D. MELIKHOV and B. STECH}

\address{Institut f\"ur Theoretische Physik, Universit\"at Heidelberg,
Philosophenweg 16, D-69120, Heidelberg, Germany} 


\maketitle
\abstracts{We calculate the form factors for weak decays of $B_{(s)}$ 
and $D_{(s)}$ mesons to light pseudoscalar and vector mesons within a
relativistic dispersion approach based on the constituent quark picture.
This approach gives the form factors as relativistic double spectral
representations in terms of the wave functions of the initial and final 
mesons. The form factors have the correct analytic properties and satisfy 
general requirements of nonperturbative QCD in the heavy quark limit. 
The effective quark masses and meson wave functions are determined 
by fitting the quark model parameters to lattice QCD results for the 
$B\to \rho$ transition form factors at large momentum transfers and to 
the measured $D\to (K,K^*)l\nu$ decay rates.
This allows us to predict numerous form factors for all kinematically
accessible $q^2$ values.}

The knowledge of the weak transition form factors of heavy mesons is crucial
for a proper extraction of the quark mixing parameters, for the analysis of
non-leptonic decays and CP violating effects and for a
search of New Physics.

Theoretical approaches for calculating these form factors are
quark models,
QCD sum rules, and lattice QCD (a detailed list of references can be found 
in \cite{ms}). Although in recent years considerable progress has been made, the
theoretical
uncertainties are still uncomfortably large. An accuracy better than 15\%
has not been attained. Moreover each of the above methods has only a limited
range of applicability: QCD sum rules are suitable for describing the 
low $q^2$ region of the form factors; lattice QCD gives good predictions for 
high $q^2$. As a result these methods do not provide for a full picture of 
the form factors and, more important, for the relations between the various 
decay channels.

Quark models do provide such relations and give the form factors in the full
$q^2$-range. However, quark models are not closely related to QCD
and therefore have input parameters which may not be of fundamental 
significance.

Clearly, a combination of various methods can be fruitful 
for obtaining reliable predictions for many decay form factors 
in their full $q^2$-ranges. To achieve this goal, one needs a general 
frame for the description of a large variety of processes. This can only be  
a suitable quark model, since only a quark model connects different 
processes through the meson soft wave functions and describes the 
full $q^2$-range of the form factors. 
This program has been implemented in our recent work \cite{ms} where 
the predictions of the quark model have been considerably improved by incorporating 
the results from lattice QCD and the available experimental data. 

\vskip0.2cm
\noindent{\bf 1. The physical picture}
\vskip0.1cm
\noindent The constituent quark picture is based on the following phenomena: 

\begin{itemize}
\item
the chiral symmetry breaking in the low-energy region which provides for 
the masses of the constituent quarks;

\item
a strong peaking of the nonperturbative meson wave functions
in terms of the quark momenta with a width of the order of
the confinement scale;  
\item
a $q\bar q$ composition of mesons in terms of the constituent quarks. 
\end{itemize}

\vskip0.2cm
\noindent{\bf 2. The formalism}
\vskip0.1cm

\noindent For the description of the transition form factors in their full
$q^2$-range and for various initial and final mesons, a fully relativistic
treatment is necessary. We make use of the dispersion formulation of the quark
model \cite{m} which guarantees the correct spectral and analytic properties
of the form factors. 

The transition form factors in the decay region are given by the relativistic
double spectral representations through the wave functions of the initial
and final mesons. 
These spectral representations obey rigorous constraints from QCD on the
structure of the long-distance corrections in the heavy quark limit:  
the form factors 
of the dispersion quark model have the correct heavy-quark expansion at
leading and next-to-leading $1/m_Q$ orders
in accordance with QCD for transitions between heavy quarks. 
For the heavy-to-light transition the dispersion quark model
satisfies the relations between the form factors
of vector, axial-vector, and tensor currents valid at small recoil. 
In the limit of the heavy-to-light transitions at small $q^2$ the form
factors obey the lowest order $1/m_Q$ and $1/E$ relations of the Large
Energy Effective Theory. 

\vskip0.2cm
\noindent{\bf 3. Parameters of the model}
\vskip0.1cm

\noindent A possible way to control quark masses and the meson wave functions
is to use the lattice results \cite{lat} for the $B\to \rho$
form factors at large $q^2$ as 'experimental' inputs. 
In \cite{mb} the $b$ and $u$ constituent quark masses
and slope parameters of the $B$, $\pi$, and $\rho$ wave functions have been
obtained through this procedure.

In \cite{ms} we have included into consideration also 
charm and strange mesons and fixed their wave functions and the effective masses 
$m_c$ and  $m_s$ by fitting the measured rates
for the decays $D\to (K, K^*)l\nu$.  

With these few inputs we gave in Ref \cite{ms} numerous predictions for the form 
factors for the $D_{(s)}$ and $B_{(s)}$ decays into light mesons which nicely 
agree at places where data are available.

\vskip0.2cm
\noindent{\bf 4. Results}
\vskip0.1cm
\noindent The main results of our analysis are as follows: 

\noindent 1. In spite of the rather different masses and properties of mesons
involved in weak transitions, all existing data on the form factors can be 
understood in the quark picture, i.e. all 
form factors can be described by the few degrees of
freedom of constituent quarks. Details of the soft wave functions are not 
crucial; only the
spatial extention  of these wave functions of order of the confinement scale
is important. In other words, only the meson radii are essential.

\noindent 
2. The calculated transition form factors are in good agreement
with the results from lattice QCD and from sum rules
in their regions of validity. The only exception is a disagreement
with the sum rules \cite{lcsr} for the $B_s\to K^*$ transition. 
This disagreement is caused by a different way of taking into account the
SU(3) violating effects when going from $B\to\rho$ to $B_s\to K^*$ and is
not related to specific details of the model.
We suspect that the sum rules \cite{lcsr} overestimate the SU(3) breaking
in the long-distance region, 
but this problem deserves further clarification. 

\noindent 
3. We have estimated the products  
of the meson weak and strong coupling constants by extrapolating 
the form factors to the meson pole. The value of each coupling constant 
can be obtained independently from the residues of several form factors. 
In all cases the values extracted from the different form factors 
agree with each other within the 5-10\% accuracy. This gives additional  
argument in support of the reliability of our estimates for the form factors. 

\section*{Acknowledgments}
It is a pleasure to thank the Organizers for creating a fruitful working 
atmosphere during this nice and interesting meeting. D.M. acknowledges 
the support of the BMBF under project 05 HT 9 HVA3.

\end{document}